\documentclass[twocolumn,pra,aps,superscriptaddress,showpacs,amsmath,amssymb,floatfix,longbibliography]{revtex4-1}
\usepackage{graphicx}
\usepackage{amssymb}
\usepackage{amsmath}
\usepackage{epsfig}
\usepackage{color}
\usepackage{mathtools}
\usepackage[colorlinks,linkcolor=blue,anchorcolor=blue,citecolor=blue,urlcolor=blue]{hyperref}
\usepackage[normalem]{ulem}

\begin{document}

\title{Proposal for quantum many-body simulation and torsional matter-wave interferometry with a levitated nanodiamond}

\author{Yue Ma}
\affiliation{Center for Quantum Information, Institute for Interdisciplinary Information Sciences, Tsinghua University, Beijing 100084, China}
\affiliation{Department of Physics, Tsinghua University, Beijing 100084, China}

\author{Thai M. Hoang}
\affiliation{Department of Physics and Astronomy, Purdue University, West Lafayette, IN 47907, USA}

\author{Ming Gong} \email{gongm@ustc.edu.cn}
\affiliation{Key Laboratory of Quantum Information, University of Science and Technology of China, Hefei, 230026, Anhui, China}
\affiliation{Synergetic Innovation Center of Quantum Information and Quantum Physics, University of Science and Technology of China, Hefei, 230026, P.R. China}

\author{Tongcang Li}\email{tcli@purdue.edu}
\affiliation{Department of Physics and Astronomy, Purdue University, West Lafayette, IN 47907, USA}
\affiliation{School of Electrical and Computer Engineering, Purdue University, West Lafayette, IN 47907, USA}
\affiliation{Purdue Quantum Center, Purdue University, West Lafayette, IN 47907, USA}
\affiliation{Birck Nanotechnology Center, Purdue University, West Lafayette, IN 47907, USA}

\author{Zhang-qi Yin}\email{yinzhangqi@mail.tsinghua.edu.cn}
\affiliation{Center for Quantum Information, Institute for Interdisciplinary Information Sciences, Tsinghua University, Beijing 100084, China}

\begin{abstract}
Hybrid spin-mechanical systems have great potentials in sensing, macroscopic quantum mechanics, and quantum information science.
In order to induce strong coupling between an electron spin and the
center-of-mass motion of a mechanical oscillator, a large magnetic gradient is usually required, which is difficult to achieve.
Here we show that strong coupling  between the electron spin of a nitrogen-vacancy (NV) center and   the torsional vibration of an optically levitated nanodiamond can be achieved in a uniform magnetic field.
  Thanks to the uniform magnetic field, multiple  spins can strongly couple to the torsional vibration at the same time. We propose to utilize this new coupling mechanism to realize the Lipkin-Meshkov-Glick (LMG) model by an ensemble of NV centers in a levitated nanodiamond. The quantum phase transition in the LMG model and finite number effects can be observed with this system. We  also propose to generate torsional superposition states and realize torsional matter-wave interferometry with spin-torsional coupling.

 \end{abstract}

\pacs{******} \keywords{***}

\maketitle

\section{introduction}

Micro and Nano-mechanical resonators in the quantum regime, based on light-matter interaction, have wide applications in quantum metrology and
quantum information science \cite{ASP2014}. It is one of the best test-beds for generating macroscopic quantum superpositions, and studying
quantum-classical boundaries \cite{PZ2012,Chen2013}. To this end, mechanical resonators need to be coupled to other systems, such as atoms \cite{Hammerer2009},
superconducting circuits \cite{Connell2010,Yin2015}, cavity modes \cite{ASP2014}, nitrogen-vacancy (NV) centers \cite{Rabl2009,Arcizet2011,Yin2015a},
etc. Among these systems, NV centers attract a great attention \cite{NVreview2013}
due to the extraordinary long coherence time (ms) even at room temperature\cite{Balasubramanian2009} as well as its high manipulation and detection
efficiency. The center-of-mass motion can even be coupled to the electron spins in magnetic field with a large gradient\cite{Rabl2009,Yin2013}; typically
this gradient should be of the order of $10^{7}$ T/m, which is difficult to achieve in stat-of-art experiments. Furthermore, this large magnetic gradient
prevents collective coupling between NV electron spin ensemble and the mechanical oscillator.

The motion of the nanoparticles can behave in a totally different way when levitated in a high quality vacuum by optical trapping\cite{Li2011, Chang2010,Romero2010}.
In this case the nanoparticles can vibrate along different directions controlled by the external optical field. In recent years this torsional vibration for a
nonspherical nanodiamond in an optical trap in vacuum was observed\cite{Hoang2016}. It was also proposed that a torsional mode can be cooled down to the ground state
by a linearly polarized cavity mode \cite{Hoang2016,Stickler2016}. In this work, we investigate the coupling between the  torsional vibration of a levitated nanodiamond and NV center electron spins
(see Fig. \ref{model}). The orientation of an NV center  will change together with the torsional vibration of the nanodiamond \cite{trappedion}, thus even in a uniform
magnetic field the energy levels of the NV electron spins can still depend strongly on its orientation as well as its displacement from the origin, which induce coupling
between the torsional vibration and the NV spin. We find that  strong coupling  can be reached with a modest uniform magnetic field (for example $0.05$ T), thus can circumvent
the technical difficulty mentioned above\cite{Rabl2009,Yin2013}.

We also propose several applications of spin-torsional coupling.
We show how to realize matter-wave interferometry, and propose to use the collective coupling between an ensemble of NV electron spins and the torsional mode to
realize the Lipkin-Meshkov-Glick (LMG) model \cite{Lipkin1965,Meshkov1965,Glick1965}. The LMG model was first introduced in nuclei physics
for phase transitions,  and has been found to be relevant to a large number of quantum systems such as  Bose-Einstein condensates in different traps \cite{Milburn1997,
Gang2009,Keeling2010, Opatrny2015}, the  Bardeen-Cooper-Schrieffer superconducting model \cite{Ortiz2005}, the radiation-matter Dicke model \cite{Latorre2005,Reslen2005,Baumann2010,Hamner2014} and
cavity QED \cite{Morrison2008}. Up to now the special case of LMG model was realized \cite{Zibold2010, Albiez05} in ultracold atoms and the corresponding transition from Rabi dynamics
to Josephson dynamics has been reported, yet the full LMG model has never been experimentally realized. We show that the quantum phase transition in the LMG model and finite number effects can be observed in our proposed system.

\section{Spin-torsional coupling}

\begin{figure}[tp]
    \includegraphics[width=4.5cm]{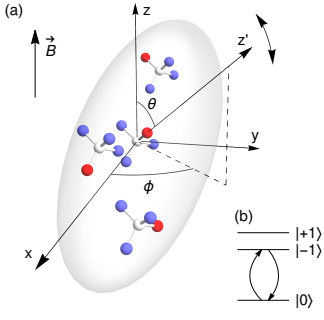}
     \caption{(Color online). (a) NV centers in a levitated diamond nanocrystal in a uniform magnetic field $\mathbf{B}$ along $z$ axis.
          We only consider NV centers in one direction ($z'$) of the four possible orientations.  The orientation of $z'$  is shown in the lab frame $oxyz$ using  polar angle $\theta$ and azimuthal angle $\phi$.(b) Energy levels for NV centers electron spins. }
  \label{model}
\end{figure}

We consider a non-spherical nanodiamond with one long axis and two short axes optically trapped in high vacuum \cite{Hoang2016,Kuhn2016} in a static uniform magnetic field (Fig. \ref{model}). The direction of the nanodiamond can be manipulated and aligned with the laser field \cite{Hoang2016,Geiselmann2013,trappedion}.
 We  consider the torsional vibration of the nanodiamond along $\theta$ direction around the polarization direction of the laser beam. The torsional Hamiltonian is $H_{tor}= \omega_\theta b^{\dagger} b$ (with natural unit $\hbar=1$). Typically $\omega_\theta$ is of the order of MHz.

We first consider a single NV center in the nanodiamond. The direction of the magnetic field is denoted as $z$ while the intrinsic quantization direction of the NV center is denoted as $z'$ (see Fig~\ref{model}).
The Hamiltonian of the NV center is $H_{NV}= DS_{z'}^2+g\mu_B {\mathbf{B}} \cdot {\mathbf S}$, where $D = 2.8$ GHz for typical NV center.
$S_{z(z')}$ is the spin-1 operator along $z(z')$ direction. If we use the eigenvectors of $S_{z'}$ to expand $H_{NV}$, and define $\Delta=-g\mu_B |{\mathbf{B}}|$,
the Hamiltonian becomes~\cite{Maclaurin2012}
\begin{equation}
    H_{\text{NV}}= \begin{pmatrix}
    D-\Delta\cos\theta & \Delta\frac{e^{-i\phi}}{\sqrt{2}}\sin\theta & 0\\
    \Delta\frac{e^{i\phi}}{\sqrt{2}}\sin\theta & 0 & \Delta\frac{e^{-i\phi}}{\sqrt{2}}\sin\theta\\
    0 & \Delta\frac{e^{i\phi}}{\sqrt{2}}\sin\theta & D+\Delta\cos\theta
    \end{pmatrix}_{z'}.
\end{equation}
If the gradient of $\mathbf{B}$ is large, the strong coupling between the translational motion of the diamond and
the spin $\mathbf{S}$  could be achieved \cite{Yin2013,Yin2015a}.

\begin{figure}[tp]
    \includegraphics[width=4.0cm]{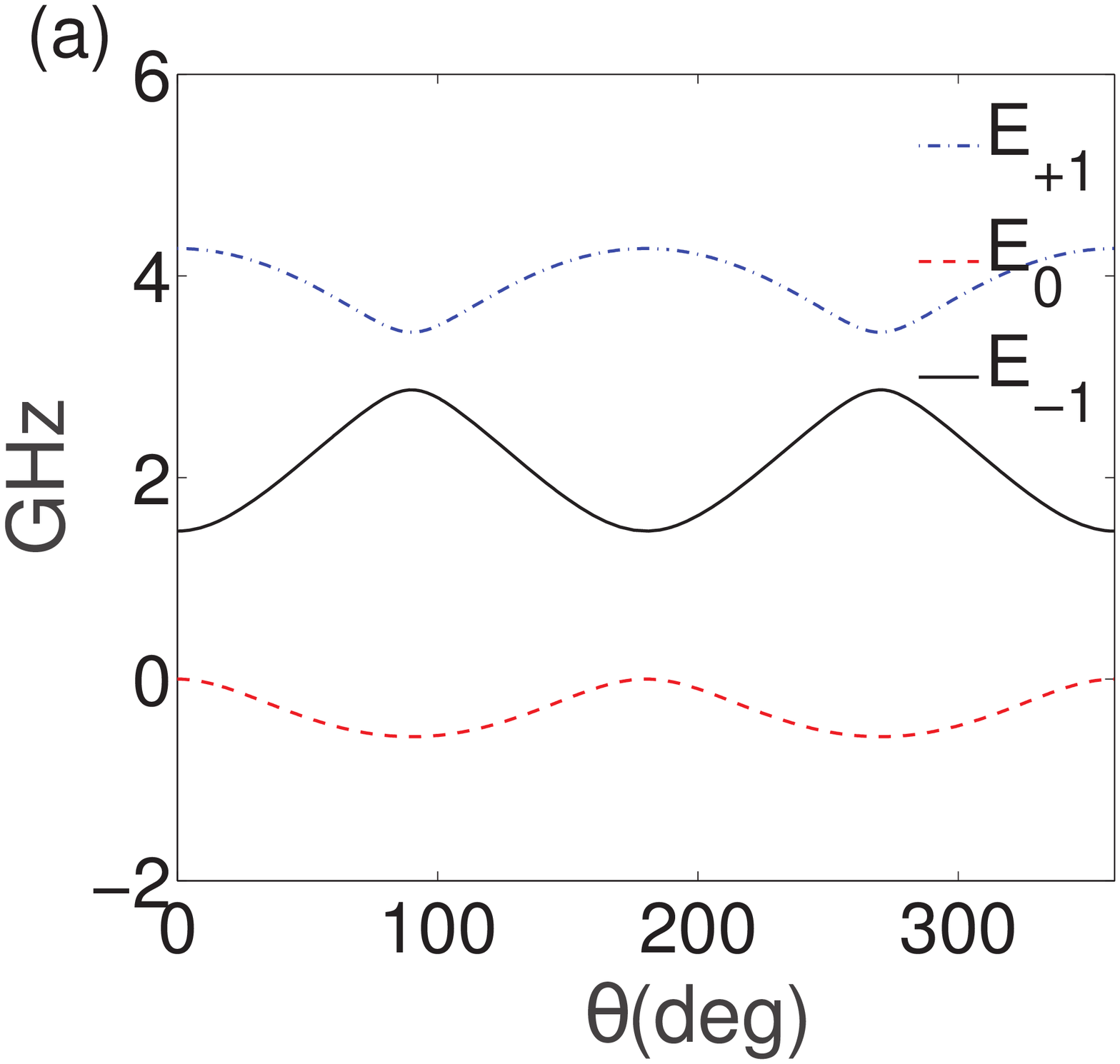}
    \includegraphics[width=4.0cm]{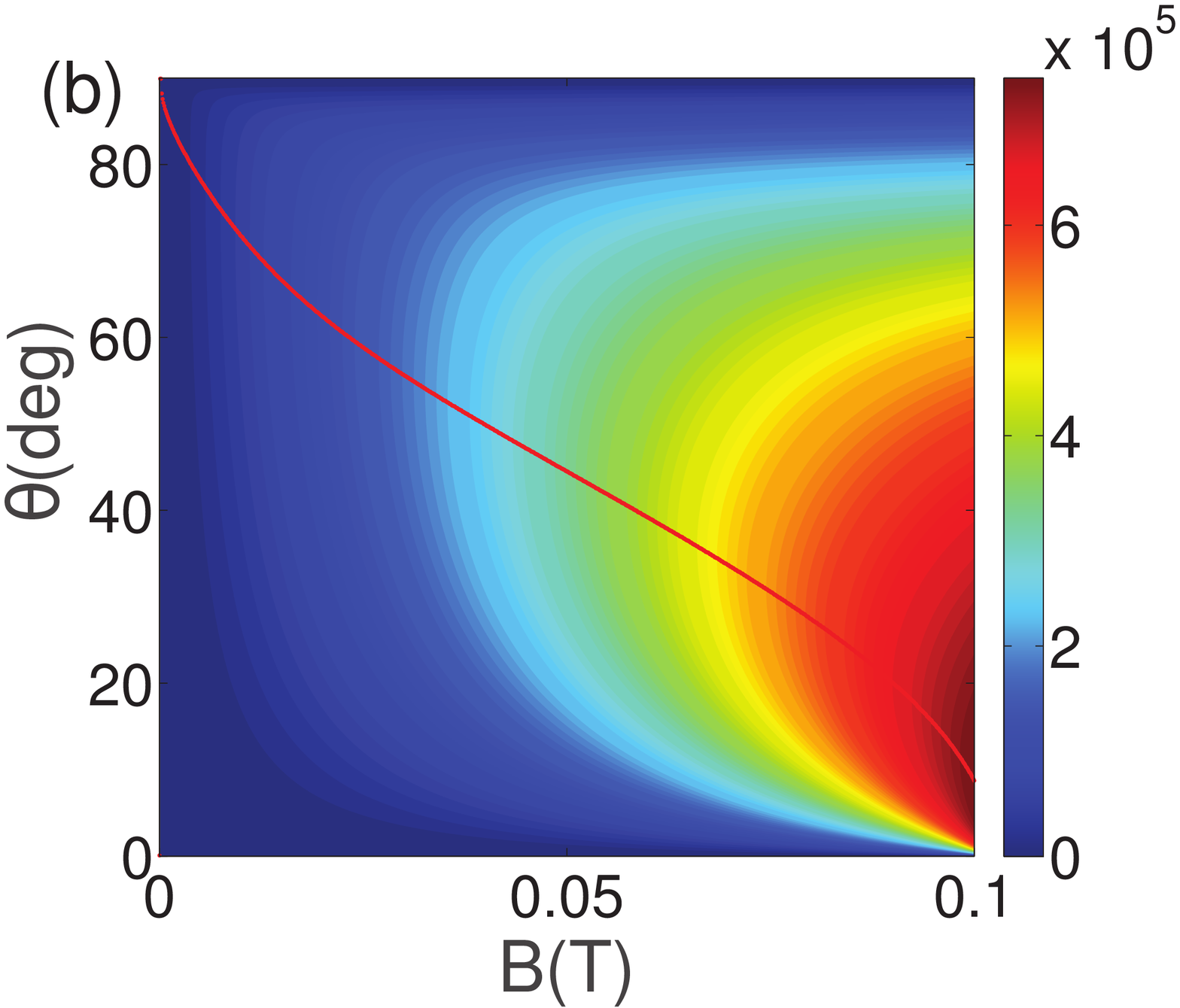}
    \caption{(Color online). (a) An example of eigenenergies of an NV center in a magnetic field as a function of the relative angle $\theta$. The magnetic field is $B=0.05\ \rm{T}$. Energy levels $E_{+1,0,-1}$ correspond to states $|S_{z''}=+1,0,-1\rangle$. (b) Spin-torsional coupling strength $g_N/2\pi$ of a nanodiamond as a function of $\theta$ and $B$. Its long(short) axis is $80$($40$) nm, its density is $3500\ \rm{kg/m^3}$, and the torsional frequency is $2.52\ \rm{MHz}$. The color shows the value of $g_N/2\pi$ in unit of Hz. The red line corresponds to the $\theta$ with the largest $g_N$ for a given $B$.}
    \label{EC}
\end{figure}

Here we suppose that $\mathbf{B}$ is homogeneous, and $\theta$ changes with the torsional motion. The angles $\theta_0$ and $\phi_0$ denote the equilibrium orientation. The eigenvalues for $H_{NV}$ is determined by the following cubic function,
\begin{equation}
    z^3 - 2 D z^2 + (D^2 - \Delta^2) z  + D \Delta^2 \sin^2(2\theta) =0,
\end{equation}
which is independent of phase $\phi$.
The calculated eigenvalues $z = E_i$ ($i=-1, 0, 1$) for a modest magnetic field as a typical example is shown in Fig~\ref{EC}(a).
We only deal with two spin states, $|S_{z''}=-1\rangle$ and $|S_{z''}=0\rangle$, where $S_{z''}$ is the spin operator in whose representation $H_{NV}(\theta_0, \phi_0)$ is diagonal.  We change the energy zero point to $E_0$ and define the energy of $|S_{z''}=-1\rangle$ as $E(\theta,\phi)=E(\theta)=E_{-1}(\theta)-E_{0}(\theta)$.

The total Hamiltonian for the NV center and the torsional oscillator reads
\begin{equation}
\label{hamiltonian1}
H=\omega_{\theta} b^{\dagger}b+\frac{E(\theta_0)}{2}\sigma_{z''}+\frac{ g_N}{2}\sigma_{z''}(b^{\dagger}+b),
\end{equation}
where $\sigma_{z''}=|S_{z''}=-1\rangle\langle S_{z''}=-1|-|S_{z''}=0\rangle\langle S_{z''}=0|$. We let $|S_{z''}=-1\rangle \equiv|-1\rangle$ and $|S_{z''}=0\rangle \equiv|0\rangle$. This is the representation that we will focus on in the rest of this paper.
The coupling between the torsional mode and NV center electron spin is
\begin{equation}
    g_N=\sqrt{\frac{1}{2I\omega_\theta}}\frac{\partial E(\theta)}{\partial \theta}\Big|_{\theta=\theta_0},
\end{equation}
where $I$ is the moment of inertia, and $\omega_\theta$ is the angular frequency of the torsional mode  \cite{Hoang2016}.
 As shown in Fig.~\ref{EC}(b), $g_N/2\pi$ could be about $300$ kHz at
$0.05$ T, which is much larger than both torsional mode decay ($<$  kHz) \cite{Hoang2016} and the NV center decay ($<$ kHz ) and dephasing ($\sim$ kHz) rates \cite{Knowles2014}. Therefore the strong coupling condition is fulfilled. In experiments the value of $g_N$ can be tuned in a wide range by controlling either the trap potential or
the external uniform magnetic field. The typical energy scales used in this paper are summarized in Table \ref{tableI}.

\section{The LMG model with levitated NV centers}

Here we show that this novel platform provides an excellent opportunity to simulate the LMG model\cite{Lipkin1965,
Meshkov1965, Glick1965}.
For the nanodiamond considered here, the mean separation between NV centers is assume to be $d > 15$~nm and the direct dipole-dipole interactions between
NV  centers  ($J_\text{d-d} < 20$~kHz \cite{Bermudez2011}) are much smaller than the spin-torsional coupling ($\sim 300$ kHz at 0.05 T). The coherence time of NV centers in nanodiamond at such low concentration could be around ms \cite{Balasubramanian2009,Knowles2014,Andrich2014}. As multiple NV centers are coupled to the same torsional mode coherently in a uniform magnetic field, the torsional mode mediating coupling between NV centers can be strong using a proper experimental scheme. To realize the LMG model, a microwave driving field with frequency $\omega_l$ and Rabi frequency $\Omega$ is added. The Hamiltonian of the system that contains multiple NV centers and the microwave drive is,
\begin{equation}
\label{hamiltonian}
H=\omega_\theta b^{\dagger}b+\frac{E_0}{2}S_{z''} + \frac{g_N}{2} S_{z''} (b^{\dagger}+b) + (\frac{\Omega}{2} e^{-i\omega_lt}S^{+} + \text{h.c.}),
\end{equation}
where $S_\alpha = \sum_{j=1}^N \sigma_{\alpha}^j$  for $\alpha =x''$, $y''$ and $z''$ are the
total spin operators for a system with $N$  NV centers in $z'$ direction. $S^+ = (S_{x''} + iS_{y''})/2$ and $S^- = (S_{x''} - iS_{y''})/2$. NV centers along other
directions (see Fig~\ref{model}) can be neglected as they have very different energy levels.
In our model, the torsional mode mediated spin flip is forbidden in the $x''-y''-z''$ coordinate. Thus  the last driving term should be included to realize
the long-range LMG model. The above time-dependent model, in which $E_0$ and $\omega_l$ are the dominante frequencies, can be transformed to the
low-frequency stationary Hamiltonian $H_{\text{eff}} = U^\dagger H U - U^\dagger i\partial_t U$ via a unitary rotation $U = e^{-i\omega_l t S_{z''}/2}$. We obtain
\begin{equation}
    H_{\text{eff}} = \omega_\theta b^{\dagger}b+\frac{\Omega}{2} S_{x''}  + \frac{E_0'}{2}S_{z''} + \frac{g_N}{2} S_{z''} (b^{\dagger}+b),
    \label{eq-Heff}
\end{equation}
where $E_0' = E_0 - \omega_l$. So the microwave driving field only affects the effective Zeeman field along the $z''$ direction. We suppose resonant driving condition
fulfilled with $E_0'=0$, in which case an isotropic LMG model can be realized.

To adiabatically eliminate the influence of the torsional mode, we need to go into a rotating frame. Let us define $H_{\text{eff}}=H_0+V$. Here $H_0 = \omega_\theta b^{\dagger}b+\frac{\Omega_0}{2} S_{x''}$ defines which rotating frame we use, where $\Omega_0$ is the frequency of the rotating frame for NV centers. In an experiment, it could be adjusted by changing the frequency of a local reference oscillator.
$V$ is treated as perturbation, which is a quite good approximation regarding that $\Omega \sim \Omega_0 \sim \omega_\theta$ is the leading term in Eq. \eqref{eq-Heff},
while all the other coefficients are much smaller than these energy scales (see Table \ref{tableI}). After the transformation the Hamiltonian $H_{\text{eff}}$ is represented by the interaction term

\begin{equation}
    V_\text{int} = \frac{g_N}{2} [\tilde{S}_{+} b^\dagger e^{i(\Omega_0 + \omega_\theta) t}+ \tilde{S}_{+} b e^{i(\Omega_0-\omega_\theta)t} +\text{h.c.}] + \frac{h_0}{2} \tilde{S}_z.
    \label{eq-Vint}
\end{equation}
where $\tilde{S}_+= (\tilde{S}_{x} +i \tilde{S}_{y})/2$, $\tilde{S}^-= (\tilde{S}_{x} -i \tilde{S}_{y})/2$ and $h_0 = \Omega - \Omega_0$. Notice that here we have adopted the notations:
$ S_{x''} = \tilde{S}_z$, $S_{y''}= -\tilde{S}_y$, and $S_{z''}= \tilde{S}_x$.

Here we consider many NV centers in a single nanodiamond.
The single spin-torsional coupling strength $g_N \propto 1/\sqrt{I \omega_\theta}$ depends on the  torsional frequency $\omega_\theta$  and the size of the nanodiamond. We suppose the concentration of NV center is kept as a constant.
As the size of nanodiamond increases, both $I$ and $N$ will increase. However, the coupling strength $g_N$ decreases. We suppose that $g_N \propto 1/\sqrt{N}$.
In the following discussion, we define $ g = \sqrt{N} g_N$ as the collective spin-torsional coupling strength.
The effective Hamiltonian can be derived using the method in Refs. [\onlinecite{James2007,Goldman2015, Bukov2015}]. In the limit that $\Omega_0\pm \omega_\theta \gg g_N, h_0$, we get
\begin{eqnarray}
    H_{\text{LMG}} &=& \frac{\lambda}{N} (\tilde{S}_x \tilde{S}_x + \tilde{S}_y \tilde{S}_y) + h\tilde{S}_z,
\end{eqnarray}
where $\lambda = \frac{g^2}{8} \cdot \frac{\omega_\theta}{\Omega_0^2 -\omega_\theta^2}$, $h = \frac{h_0}{2} + \frac{\Omega_0}{\Omega_0^2 - \omega_\theta^2}\cdot (b^\dagger b + \frac{1}{2}) {g^2/ 2N}$.

We mainly focus on the condition of ferromagnetic coupling ($\lambda < 0$) when $\omega_\theta > \Omega_0$. The effective Zeeman field is consisted of two parts: the contribution from the external Zeeman field inherent from the original Hamiltonian and the contribution from phonons. Notice that the second part is  due to the fact that the interaction $V_\text{int}$ breaks the time-reversal symmetry. This term will  become unimportant if the total number of NV centers $N$ is large enough. The first term $h_0$ will play primary role
for the phase transition near torsional ground state or when $N$ is much larger than the torsional thermal phonon number $\langle b^\dagger b\rangle$. For $h \ll \Omega_0$ the change of $h$ will not significantly change the coupling strength $\lambda$, thus these two parameters may be treated as
independent parameters. In the above calculations, we focused on the resonant excitation ($E_0' = 0$). Thus the two orthogonal directions
($x$ and $y$) are equivalent. When $E_0' \ne 0$, the equivalence between these two directions is broken and the asymmetric LMG model may also be realized. If there is no  driving field, the spin-flipping is forbidden thus only long-range classical Ising model can be realized [\onlinecite{Wei2015}],
which may also support phase transition belonging to a totally different universal class due to the fact that a $d$ (here $d=1$) dimensional quantum spin model is equivalent
to the $(d+1)$ dimensional classical model. This system also has the potential to study antiferromagnetism \cite{Cheng2016}.

\begin{table}
      \caption{Typical parameters. All the frequencies are in unit of $2\pi\times$ MHz.}
      \begin{tabular*}{3.0in}{ c | c | c | c | c | c | c | c}
      \hline
          $(E_0, \omega_l)$  & Size  & $\omega_\theta$  & $g_N$  & $\Omega_0$   & $g$  & $\lambda$ & $|h_0|$ \\ \hline
          $\sim 2000 $          & (80, 40) nm    &  2.52  &  0.1  & 1.0  & 0.42  & 0.01 & $\sim 0.01$ \\ \hline
      \end{tabular*}
      \label{tableI}
\end{table}

\begin{figure}[tp]
    \centering
    \includegraphics[width=2.8in]{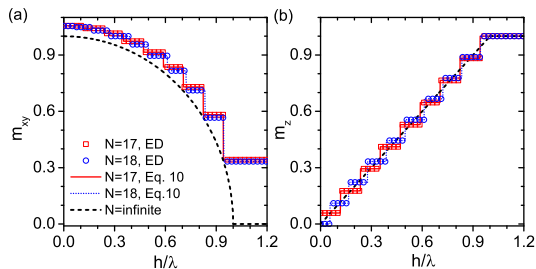}
    \caption{(Color online). Phase transition of multiple NV spins in a levitated nanodiamond in a magnetic field. A small number of NV centers is sufficient to see the signature of phase transition. (a) $m_{xy}$ and (b) $m_{z}$ are the in-plane and out-of-plane polarizations, respectively.
    The square and circle symbols represent results obtained from exact diagonalization of the Hamiltonian, while solid and dotted lines are the analytical results in Eq. \ref{eq-10}. The dashed line is the exact result in the thermodynamic limit.
}
        \label{fig-mxyz}
\end{figure}

In the thermodynamic limit, the LMG model hosts phase transition at $h = \lambda$.  In realistic systems, the number of NV centers in a diamond nanocrystal is always finite and can be controlled by doping. For a 80~nm-diameter diamond, the total number of NV color centers can be in the range of $N \sim 0 - 100$, which is sufficient to observe the phase transitions
\cite{Neukirch2015,Hoang2016a}. In this model the order parameters are defined as $m_{xy}=\frac{2}{N} \sqrt{\langle\tilde{S}_{x}^2 + \tilde{S}_y^2\rangle}$ for
the in-plane polarization, and $m_z=\frac{2}{N} \sqrt{\langle \tilde{S}_z^2\rangle}$ for the out-of-plane polarization\cite{Botet1983, Dusuel2005, Vidal2004}. For a finite system, we find $m_z = 1$ when $|h/\lambda|\geq 1$, and otherwise,
\begin{equation}
    m_z =
      \begin{cases}
          \frac{1}{N} + [\frac{hN}{2 \lambda}] \frac{2}{N},       & \text{ $N$ is odd} \\
          [\frac{hN}{2\lambda} + \frac{1}{2} ] \frac{2}{N},     & \text{ $N$ is even} \\
      \end{cases}
      \label{eq-10}
\end{equation}
where $[x]$ takes the integer part of $x$.  The in-plane polarization can be determined via
$m_{xy}=\sqrt{1+ \frac{2}{ N}-m_z^2 }$. We see that the discontinuous jump of $m_z$ and $m_{xy}$ can be observed at
$\frac{h }{\lambda} = \frac{2k }{ N}$ for $k = 1, 2, 3, \cdots, [\frac{N }{ 2}]$ when $N$ is odd and
$\frac{h }{\lambda} = \frac{2k-1 }{ N}$ for $k = 1, 2, 3, \cdots, [\frac{N }{ 2}]$ when $N$ is even. The jump of these
quantities arises from the quantization of the spin. These two polylines will collapse to the well-known continue
limit, $m_{xy} = \sqrt{1-(h/\lambda)^2}$ and $m_z = \frac{h}{ \lambda}$ for $|h/\lambda| < 1$, when $N\rightarrow \infty$.

Both analytical results and  numerical results for $m_{xy}$ and $m_z$ (see Fig. \ref{fig-mxyz}) suggest that the
strong signature of quantum phase transition can be observed even in a small system.
When $|h/\lambda| > 1$, the out-of-plane polarization $m_z$ exactly equals to one, while the in-polarization can
still be finite ($m_{xy}=\sqrt{\frac{2}{ N}}\approx 0.3$ when $N = 18$). This is
different from the proposal for classical phase transition in Ref. \onlinecite{Wei2015}, where an experimentally
observable effect requires an extraordinary large number of NV centers.

To observe the phase transition, we can measure $m_z$ as a function of $h/\lambda$. We prepare the NV centers in
ground state, then adiabatically tune the parameter $h/\lambda$. The exact degeneracy at the jumping point should
be removed by a modest in-plane Zeeman field. Experimentally, we measure the population of the spin in the intrinsic
quantization direction $z'$. We can first apply a microwave pulse that  rotates the state $\tilde{S}_z$ to the state
$S_{z'}$, and then measure the spin state in the intrinsic frame.

\section{Schr\"{o}dinger's cat state and torsional matter-wave interferometry}

Schr\"{o}dinger's cat state is generally considered as an entangled state between a microscopic  quantum system and a macroscopic system. It can be prepared with an optically levitated nanodiamond   using the coupling between the center-of-mass motion and the electron spin with a strong magnetic gradient ~\cite{Yin2013}. Here we show how to realize Schr\"{o}dinger's cat state with torsional motion and spin in a uniform magnetic field.

First we need to cool the torsional motion near ground state $|0\rangle_\theta$ by sideband cooling \cite{Hoang2016,Marquardt2007,Wilson2007}. Then we adiabatically lower the trapping frequency from $\omega_\theta$ to $\omega_{\theta}'$ so that $|0\rangle_{\omega_\theta}$ evolves to a new vacuum state $|0\rangle_{\omega_\theta'}$. The spin is initialized to $(|0\rangle+|-1\rangle)/\sqrt{2}$.
From $t=0$, the system evolves under the Hamiltonian~\eqref{hamiltonian1} where $\omega_\theta$ is replaced by $\omega_{\theta}'$.  We find that the system splits into two torsional oscillations centered at slightly different orientations and coupled with the two spin states respectively (see appendix \ref{cat}). At  time $t_0=\pi/\omega_{\theta}'$, the separation of orientation is the largest. The state at this time is
\begin{equation}
\label{eq:state}
\begin{aligned}
|\psi(t_0)\rangle=&\frac{1}{\sqrt{2}}\Big(e^{-\frac{iE(\theta_0)t_0}{2}}|-1\rangle\otimes e^{-\frac{g_N}{\omega_{\theta}'}(b^{\dagger}-b)}|0\rangle_{\omega_\theta'}\\
&+e^{\frac{iE(\theta_0)t_0}{2}}|0\rangle\otimes e^{\frac{g_N}{\omega_{\theta}'}(b^{\dagger}-b)}|0\rangle_{\omega_\theta'}\Big),
\end{aligned}
\end{equation}
which is the Schr\"{o}dinger's cat state.
For convenience we define $\beta=g_N/\omega_{\theta}'$. With the displacement operator $\hat{D}(\beta)=\exp(\beta b^{\dagger}-\beta^{*}b)$, we get $\hat{D}(\beta)|0\rangle_{\omega_\theta'}=|\beta\rangle_{\omega_\theta'}$, and $\hat{D}(-\beta)|0\rangle_{\omega_\theta'}=|-\beta\rangle_{\omega_\theta'}$.

\begin{figure}[tp]
     \includegraphics[width=.21\textwidth]{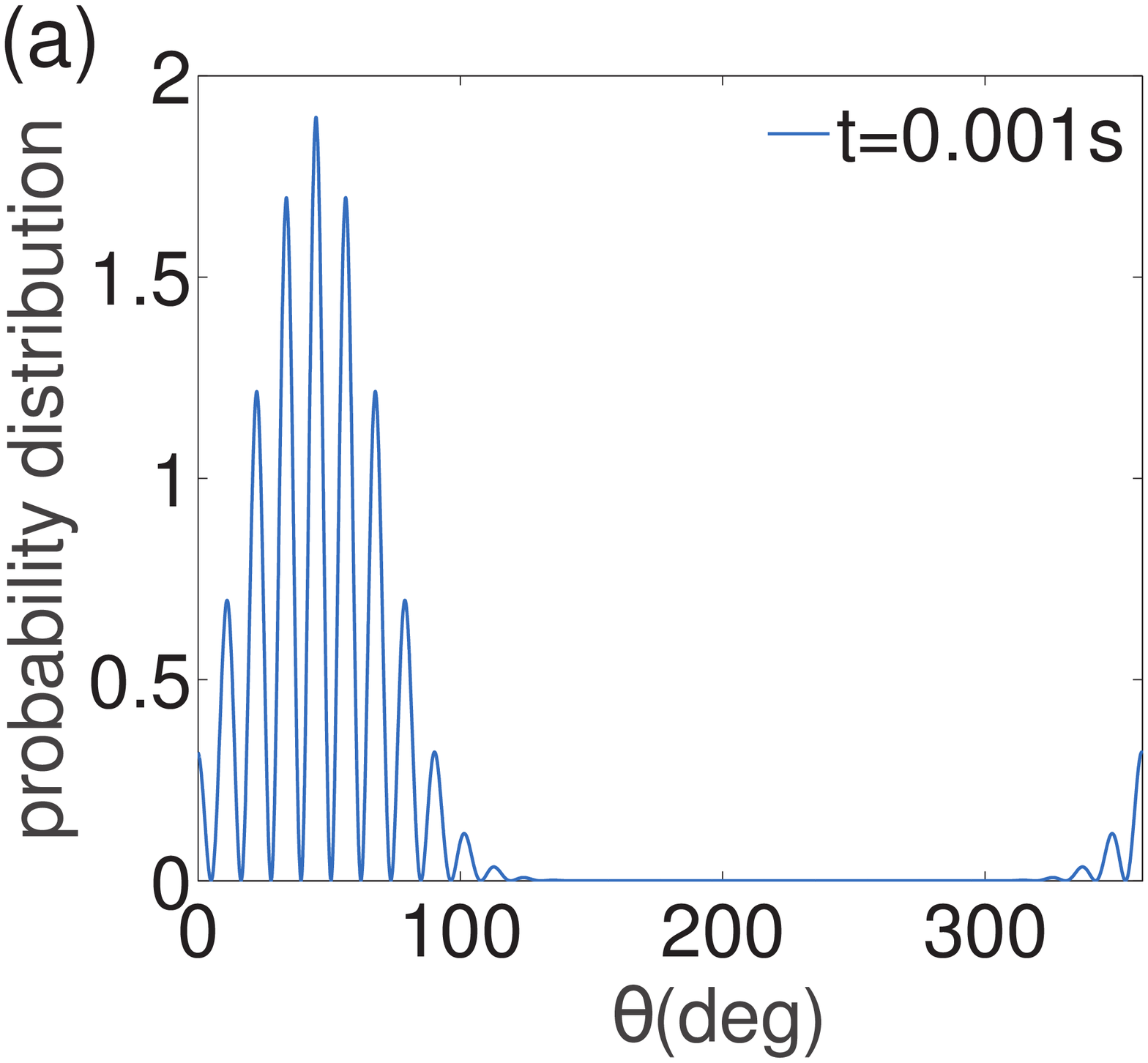}
     \includegraphics[width=.21\textwidth]{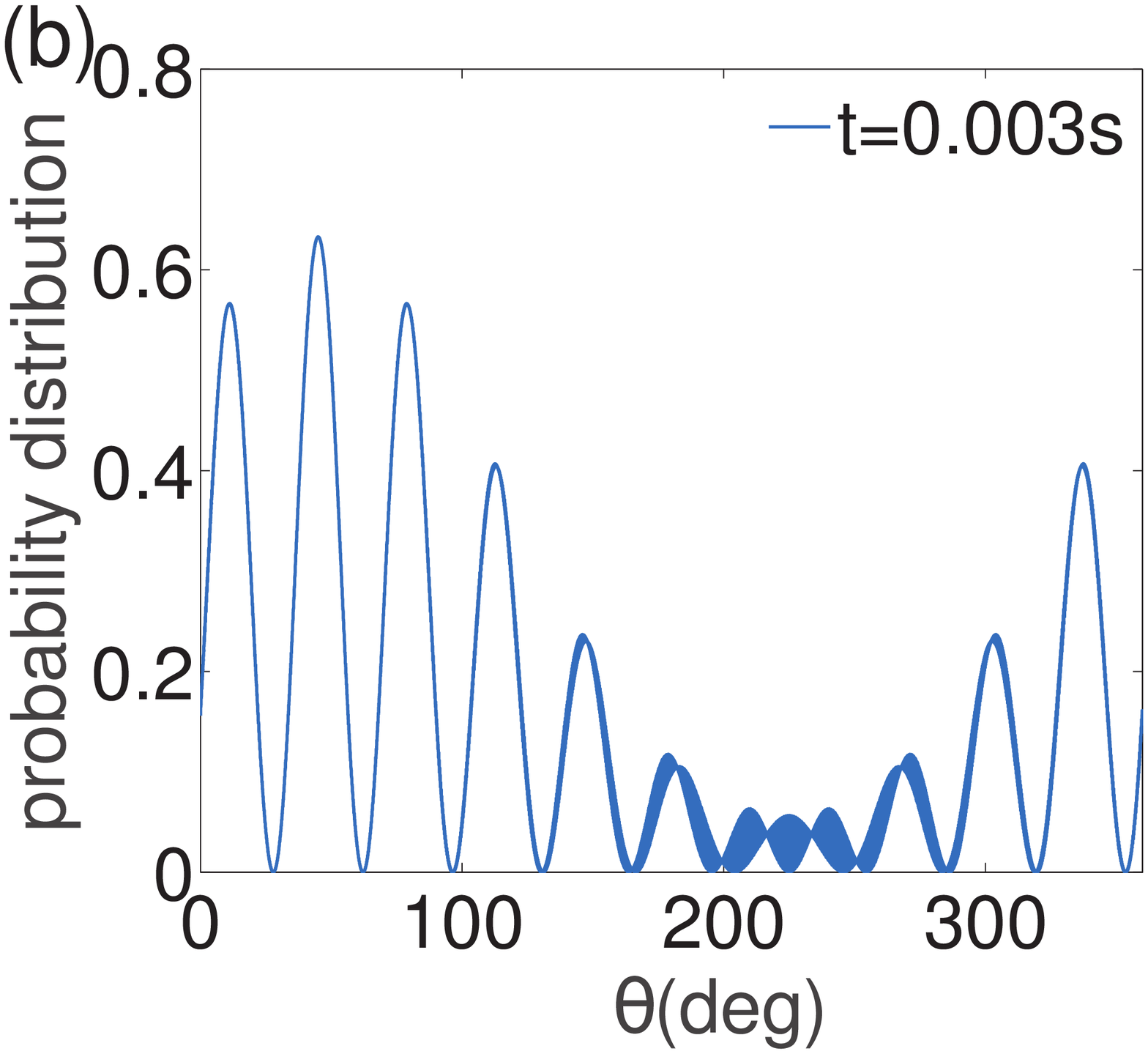}
    \caption{(Color online). Interference fringes of a nanodiamond with long axis $80\ \rm{nm}$ and short axis $40\ \rm{nm}$. The external magnetic field is $0.05\ \rm{T}$. The coupling strength is $g_N=2\pi\times331\ \rm{kHz}$ at the equilibrium orientation $\theta_0=\pi/4$. The trapping frequency is $\omega_{\theta'}=2\pi\times50\ \rm{kHz}$. (a) Fringes at $t=0.001\ \rm{s}$. The wavefunction occupies half of the circle, and the fringes are clear. (b) Fringes at $t=0.003\ \rm{s}$.  The spacing between two fringes ($ 30^{\circ}$) is large enough to be detected experimentally.
    }\label{fringe}
\end{figure}

We can use matter-wave interferometry to verify the creation of  the Schr\"{o}dinger's cat state. First, we impose pulses to disentangle the spin and the oscillation
\cite{Yin2013}. The torsional motion state becomes
$|\varphi'(t_0)\rangle=\frac{1}{\sqrt{2}}(|\beta\rangle_{\omega_\theta'}+|-\beta\rangle_{\omega_\theta'})$.
Then we turn off the magnetic field and the optical tweezer. The spin state $|0\rangle$ tends to rotate to the lowest energy state adiabatically similar as the trend in Fig.~\ref{EC}(a) under the influence of the earth magnetic field. Its timescale is larger than $0.1\ \rm{s}$, slow enough to be neglected.
Suppose $zn$ is the direction perpendicular  to $oz$ and $oz'$ in Fig.~\ref{model}(a).
The Hamiltonian for the system is $H_{free}=L_{zn}^2/2I$, where $L_{zn}$ is the orbital angular momentum operator along $zn$.
The orientation of the nanodiamond evolves freely and creates an interference pattern in $\theta$ direction (see appendix~\ref{interference}).

Fig.~\ref{fringe} shows  examples of the calculated interference fringe. When the evolutionary time is not very long, the longer the time is, the wider the fringe will be. But when the time is too long,  the wavefunction spreads over a full $2\pi$ angle, the interference fringe becomes very complex. We only need to focus on the region without the complicate pattern.

\section{Conclusion}

We discuss the strong coupling mechanism between the torsional motion of a nanodiamond and the spin of built-in NV centers under a homogeneous magnetic field.
This novel system can used for simulating LMG model and the
many-body phase transitions. Strong evidence for this phase transition can even be observed for a small nanodiamond containing only a few tens of NV centers.
 The system may also be used to realize Schr\"{o}dinger cat state and the corresponding torsional matter-wave interferometry.

\begin{acknowledgments}
Z.Q.Y. is supported by National Natural Science Foundation of China NO. 61435007, 11574176, and the Joint Foundation of Ministry of Education of China (6141A02011604).
T.L. is supported by National Science Foundation under Grant No. 1555035-PHY. 
M.G. was supported by the National Youth Thousand Talents Program (Grant
No. KJ2030000001), USTC start-up funding (Grant No.
KY2030000053), and CUHK RGC (Grant No. 401113).
We thank F. Robicheaux for helpful discussions.
M.G. thanks S.Y. Zhang for numerical assistant.
\end{acknowledgments}

\appendix

\section{Generation of Schr{\"{o}}dinger's cat state}\label{cat}
In this part, we provide a derivation of the Eq.~\eqref{eq:state}.
Suppose the system evolves under the Hamiltonian

\begin{equation}
H'=\omega'_{\theta}b^{\dagger}b+\frac{E(\theta_0)}{2}\sigma_{z''}+\frac{g_N}{2}\sigma_{z''}(b^{\dagger}+b)
\end{equation}
What we would like to do is calculating $e^{-iH't}(|0\rangle+|-1\rangle)/\sqrt{2}\otimes|0\rangle_{\omega'_{\theta}}$, but this cannot be done directly. Instead, it is better to find a way to consider the effect of the parts in $H'$ separately. In a rotating frame, $H'$ becomes

\begin{equation}
H_I'=\frac{g_N}{2}\sigma_{z''}(b^{\dagger}e^{i\omega'_{\theta}t}+be^{-i\omega'_{\theta}}t)
\end{equation}

$H_I'$ is time-dependent, and its corresponding time evolution operator is $U_I=\mathcal{T}e^{-i\int H_I'(t)dt}$, $\mathcal{T}$ is the time-ordering operator. $U_I$ can be expanded by Magnus expansion, $U_I=e^{\Sigma_{k=1}^{\infty}\Omega_k(t)}$. The first order is

\begin{equation}
\begin{aligned}
\Omega_1&=-i\int_{0}^{t}H_I'(t_1)dt_1\\
&=\frac{g_N\sigma_{z''}}{2\omega'_{\theta}}\Big((e^{-i\omega'_{\theta}t}-1)b-(e^{i\omega'_{\theta}t}-1)b^{\dagger}\Big)
\end{aligned}
\end{equation}

The second order is

\begin{equation}
\begin{aligned}
\Omega_2&=\frac{1}{2}(-i)^2\int_0^t\int_0^{t_1}[H_I'(t_1),H_I'(t_2)]dt_1dt_2\\
&=i\frac{g_N^2}{4}\Big(\frac{t}{\omega'_{\theta}}-\frac{\sin(\omega'_{\theta}t)}{\omega'^2_{\theta}}\Big)
\end{aligned}
\end{equation}

As $\Omega_2$ does not contain any operators, it is just a global phase of the state and can be neglected in our situation. Moreover, $\Omega_3=\Omega_4=\cdots=0$. So $U_I=e^{\beta(t)b-\beta(t)^{*}b^{\dagger}}$, where $\beta(t)=\frac{g_N\sigma_{z''}}{2\omega'_{\theta}}\Big(e^{-i\omega'_{\theta}t}-1\Big)$.

After we return to the original frame, the quantum state at time $t$ can be expressed as

\begin{equation}
\label{state}
\begin{aligned}
|\psi(t)\rangle=&e^{-i\omega'_{\theta}b^{\dagger}bt}e^{-\frac{iE(\theta_0)t}{2}\sigma_{z''}}e^{\frac{\sigma_{z''}g_N}{2\omega'_{\theta}}\Big((e^{-i\omega'_{\theta}
t}-1)b-(e^{i\omega'_{\theta}t}-1)b^{\dagger}\Big)}\\
&\frac{1}{\sqrt{2}}(|-1\rangle+|0\rangle)\otimes|0\rangle_{\omega'_{\theta}}\\
=&e^{-i\omega'_{\theta}b^{\dagger}bt}e^{-\frac{iE(\theta_0)t}{2}\sigma_{z''}}\\
&\frac{1}{\sqrt{2}}\Big(|-1\rangle\otimes e^{\frac{g_N}{2\omega'_{\theta}}\big((e^{-i\omega'_{\theta}t}-1)b-(e^{i\omega'_{\theta}t}-1)b^{\dagger}\big)}|0\rangle_{\omega'_{\theta}}\\
&+|0\rangle\otimes e^{-\frac{g_N}{2\omega'_{\theta}}\big((e^{-i\omega'_{\theta}t}-1)b-(e^{i\omega'_{\theta}t}-1)b^{\dagger}\big)}|0\rangle_{\omega'_{\theta}}\Big)\\
=&e^{-i\omega'_{\theta}b^{\dagger}bt}\frac{1}{\sqrt{2}}\Big(e^{-\frac{iE(\theta_0)t}{2}}|-1\rangle\\
&\otimes e^{\frac{g_N}{2\omega'_{\theta}}\big((e^{-i\omega'_{\theta}t}-1)b-(e^{i\omega'_{\theta}t}-1)b^{\dagger}\big)}|0\rangle_{\omega'_{\theta}}\\
&+e^{\frac{iE(\theta_0)t}{2}}|0\rangle\otimes e^{-\frac{g_N}{2\omega'_{\theta}}\big((e^{-i\omega'_{\theta}t}-1)b-(e^{i\omega'_{\theta}t}-1)b^{\dagger}\big)}|0\rangle_{\omega'_{\theta}}\Big)
\end{aligned}
\end{equation}

While $\exp\Big(\frac{g_N}{2\omega'_{\theta}}\big((e^{-i\omega'_{\theta}t}-1)b-(e^{i\omega'_{\theta}t}-1)b^{\dagger}\big)\Big)$ and $\exp\Big(-\frac{g_N}{2\omega'_{\theta}}\big((e^{-i\omega'_{\theta}t}-1)b-(e^{i\omega'_{\theta}t}-1)b^{\dagger}\big)\Big)$ are likely displacement operators, the effect of $e^{-i\omega'_{\theta}b^{\dagger}bt}$ cannot be seen directly. In fact, if $e^{-i\omega'_{\theta}b^{\dagger}bt}$ first acts on the state $|0\rangle_{\omega'_{\theta}}$, it will become just a time-dependent phase factor. So we now develop how to ``interchange" the first and second factor.

Suppose $e^Xe^Y=Ke^X$, $K$ is the operator we need to find out.

\begin{equation}
\label{lemma}
\begin{aligned}
K=&e^Xe^Ye^{-X}\\
=&e^X(1+Y+\frac{1}{2!}Y^2+\frac{1}{3!}Y^3+\cdots)e^{-X}\\
=&1+e^XYe^{-X}+\frac{1}{2!}e^XYe^{-X}e^XYe^{-X}\\
&+\frac{1}{3!}e^XYe^{-X}e^XYe^{-X}e^XYe^{-X}+\cdots\\
=&1+e^XYe^{-X}+\frac{1}{2!}(e^XYe^{-X})^2+\frac{1}{3!}(e^XYe^{-X})^3+\cdots\\
=&e^{e^XYe^{-X}}\\
=&e^{Y+[X,Y]+\frac{1}{2!}[X,[X,Y]]+\frac{1}{3!}[X,[X,[X,Y]]]+\cdots}
\end{aligned}
\end{equation}

In the last equality, we use the Baker-Hausdorff lemma~\cite{Sakurai2011}. First we calculate $e^{-i\omega'_{\theta}b^{\dagger}bt}\exp\Big(\frac{g_N}{2\omega'_{\theta}}\big((e^{-i\omega'_{\theta}t}-1)b
-(e^{i\omega'_{\theta}t}-1)b^{\dagger}\big)\Big)|0\rangle_{\omega'_{\theta}}$. We let $X=-i\omega'_{\theta}b^{\dagger}bt$ and $Y=\frac{g_N}{2\omega'_{\theta}}\big((e^{-i\omega'_{\theta}t}-1)b
-(e^{i\omega'_{\theta}t}-1)b^{\dagger}\big)$ to use Equation~\eqref{lemma}. Then we can get

\begin{equation}
\begin{aligned}
&e^{-i\omega'_{\theta}b^{\dagger}bt}e^{\frac{g_N}{2\omega'_{\theta}}\big((e^{-i\omega'_{\theta}t}-1)b
-(e^{i\omega'_{\theta}t}-1)b^{\dagger}\big)}|0\rangle_{\omega'_{\theta}}\\
&=e^{\frac{g_N}{2\omega'_{\theta}}\big((1-e^{i\omega'_{\theta}t})b-(1-e^{-i\omega'_{\theta}t})b^{\dagger}\big)}
e^{-i\omega'_{\theta}b^{\dagger}bt}|0\rangle_{\omega'_{\theta}}
\end{aligned}
\end{equation}

The second term in Eq.~\eqref{state} can be dealt with under similar process. So finally the state in Eq.~\eqref{state} is changed to
\begin{equation}
\begin{aligned}
|\psi(t)\rangle=&\frac{1}{\sqrt{2}}\Big(e^{-\frac{iE(\theta_0)t}{2}}|-1\rangle\\
&\otimes e^{\frac{g_N}{2\omega'_{\theta}}\big((1-e^{i\omega'_{\theta}t})b-(1-e^{-i\omega'_{\theta}t})b^{\dagger}\big)}
e^{-i\omega'_{\theta}b^{\dagger}bt}|0\rangle_{\omega'_{\theta}}\\
&+e^{\frac{iE(\theta_0)t}{2}}|0\rangle\otimes e^{-\frac{g_N}{2\omega'_{\theta}}\big((1-e^{i\omega'_{\theta}t})b-(1-e^{-i\omega'_{\theta}t})b^{\dagger}\big)}\\
&e^{-i\omega'_{\theta}b^{\dagger}bt}|0\rangle_{\omega'_{\theta}}\Big)\\
=&\frac{1}{\sqrt{2}}\Big(e^{-\frac{iE(\theta_0)t}{2}}|-1\rangle\\
&\otimes e^{\frac{g_N}{2\omega'_{\theta}}\big((1-e^{i\omega'_{\theta}t})b-(1-e^{-i\omega'_{\theta}t})b^{\dagger}\big)}
e^{-i\omega'_{\theta}tn}|0\rangle_{\omega'_{\theta}}\\
&+e^{\frac{iE(\theta_0)t}{2}}|0\rangle\otimes e^{-\frac{g_N}{2\omega'_{\theta}}\big((1-e^{i\omega'_{\theta}t})b-(1-e^{-i\omega'_{\theta}t})b^{\dagger}\big)}\\
&e^{-i\omega'_{\theta}tn}|0\rangle_{\omega'_{\theta}}\Big)
\end{aligned}
\end{equation}

From the definition of displacement operator $\hat{D}(\beta)=e^{\beta b^{\dagger}-\beta^{*}b}$ which displaces $b$ to $b+\beta$ and $b^{\dagger}$ to $b^{\dagger}+\beta^{*}$, we can see clearly that in our system when $t_0=\frac{\pi}{\omega'_{\theta}}$, the displacement of the equilibrium orientation is the maximum. Thus we get
\begin{equation}
\begin{aligned}
|\psi(t_0)\rangle=&\frac{1}{\sqrt{2}}\Big(e^{-\frac{iE(\theta_0)t_0}{2}}|-1\rangle\otimes e^{-\frac{g_N}{\omega'_{\theta}}(b^{\dagger}-b)}|0\rangle_{\omega'_{\theta}}\\
&+e^{\frac{iE(\theta_0)t_0}{2}}|0\rangle\otimes e^{\frac{g_N}{\omega'_{\theta}}(b^{\dagger}-b)}|0\rangle_{\omega'_{\theta}}\Big).
\end{aligned}
\end{equation}
This is Eq.~\eqref{eq:state} in the main text.

\section{Torsional matter waver interference}\label{interference}
In this part, we discuss how to get the interference fringes as shown in Fig.~\ref{fringe} in the main text. Initially, the wavefunction expressed by the orientation $\theta$ is $\psi_0(\theta,t=0)=\Big(\varphi_{+}(\theta)+\varphi_{-}(\theta)\Big)/\sqrt{2}$.

\begin{equation}
\begin{aligned}
\varphi_{+}(\theta)&=\langle\theta|\beta\rangle_0\\
&=\sqrt[4]{\frac{I\omega'_{\theta}}{\pi}}e^{-\frac{1}{2}I\omega'_{\theta}\Big(\theta-(\theta_0+\sqrt{\frac
{2}{I\omega'_{\theta}}}\frac{g_N}{\omega'_{\theta}})\Big)^2}\\
\varphi_{-}(\theta)&=\langle\theta|-\beta\rangle_0\\
&=\sqrt[4]{\frac{I\omega'_{\theta}}{\pi}}e^{-\frac{1}{2}I\omega'_{\theta}\Big(\theta-(\theta_0-\sqrt{\frac
{2}{I\omega'_{\theta}}}\frac{g_N}{\omega'_{\theta}})\Big)^2}
\end{aligned}
\end{equation}

$\theta_0$ is the original equilibrium orientation. Normally the normalization of a Gaussian distribution requires the argument to range from $-\infty$ to $+\infty$. But here $\theta$ can only take values between $0$ and $2\pi$. However, for a Gaussian distribution with mean value $\mu$ and standard deviation $\sigma$, the probability is almost 0 if the argument is out of the range $(\mu-3\sigma,\mu+3\sigma)$. So the lower and upper limits can be extended to $-\infty$ and $+\infty$, respectively. We will show later that this requirement is very easy to be fulfilled in experiment.

As in the standard quantum solutions, if we want to find the time evolution of a wave function, we should decompose it to the linear superposition of the eigenfunctions of the Hamiltonian, because we know the time evolution of the Hamiltonian's eigenfunctions is just the original state multiplied by a factor $e^{-iE_nt}$, $E_n$ is the eigenvalue. This applies to all time-independent Hamiltonian, so it is also suitable for our $H_{free}$ here.

We expand $\varphi_{+}(\theta)$ with the eigenfunctions of $L_{zn}$ in spherical coordinates, $\frac{1}{\sqrt{2\pi}}e^{im\theta}$ $(m=0,\pm1,\pm2,\cdots)$.
\begin{equation}
\varphi_{+}(\theta)=\sum_{m=-\infty}^{+\infty}A_{m}^{+}\frac{1}{\sqrt{2\pi}}e^{im\theta}
\end{equation}
According to the orthonormal property of $e^{im\theta}$, we have
\begin{equation}
\begin{aligned}
A_{m}^{+}=&\int_{0}^{2\pi}\varphi_{+}(\theta)\frac{1}{\sqrt{2\pi}}e^{-im\theta}d\theta\\
=&\sqrt[4]{\frac{I\omega'_{\theta}}{4\pi^3}}\int_{-\infty}^{+\infty}e^{-\frac{I\omega'_{\theta}}{2}\Big(\theta+\big(im
\frac{1}{I\omega'_{\theta}}-(\theta_0+\sqrt{\frac{2}{I\omega'_{\theta}}}\frac{g_N}{\omega'_{\theta}})\big)\Big)^2}\\
&\times e^{-\frac{m^2}{2I\omega'_{\theta}}}e^{-im\Big(\theta_0+\sqrt{\frac{2}{I\omega'_{\theta}}}\frac{g_N}{\omega'_{\theta}}\Big)}d\theta
\end{aligned}
\end{equation}

\begin{figure*}[hbtp]
\centering
    \includegraphics[width=.3\textwidth]{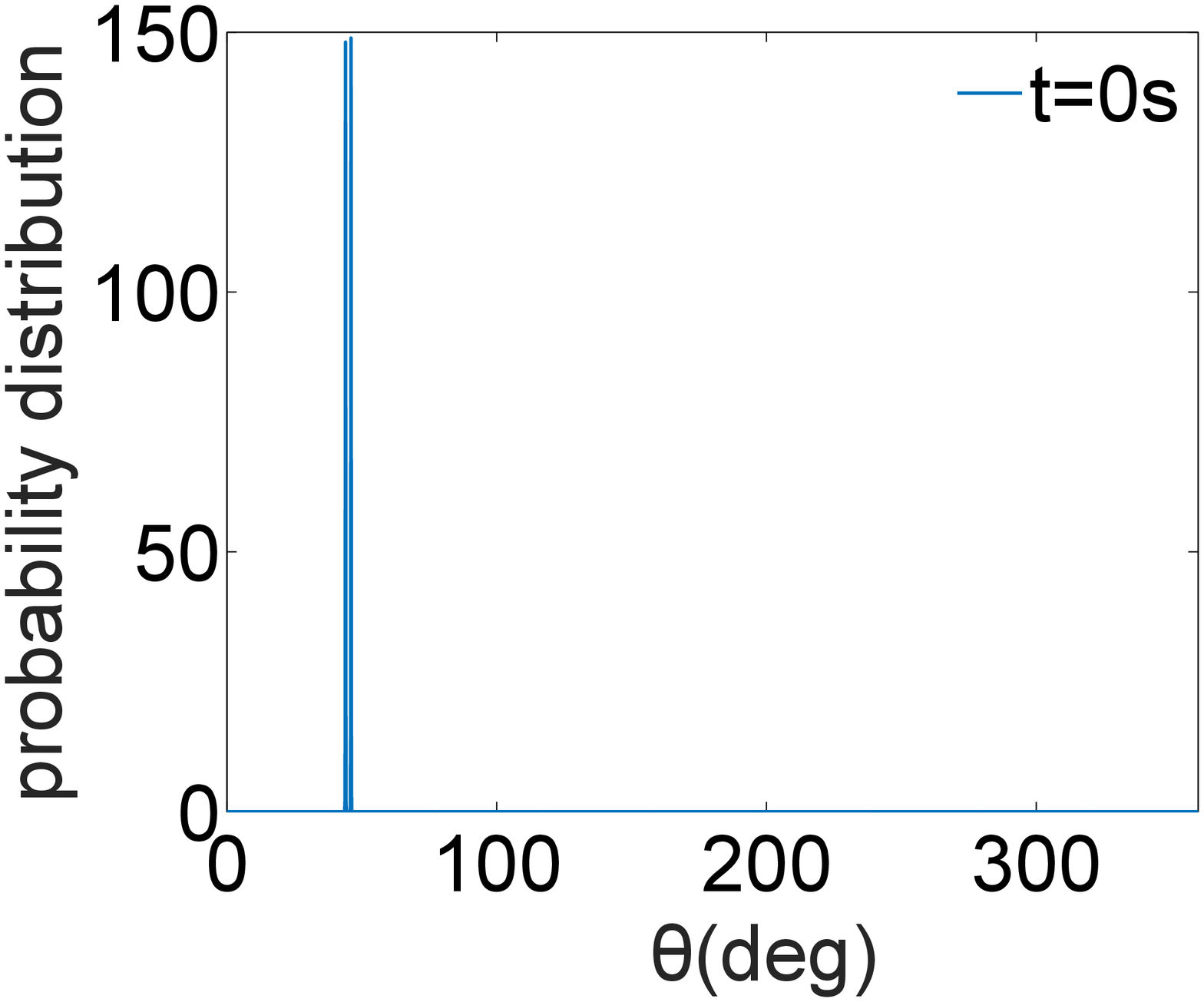}
    \includegraphics[width=.3\textwidth]{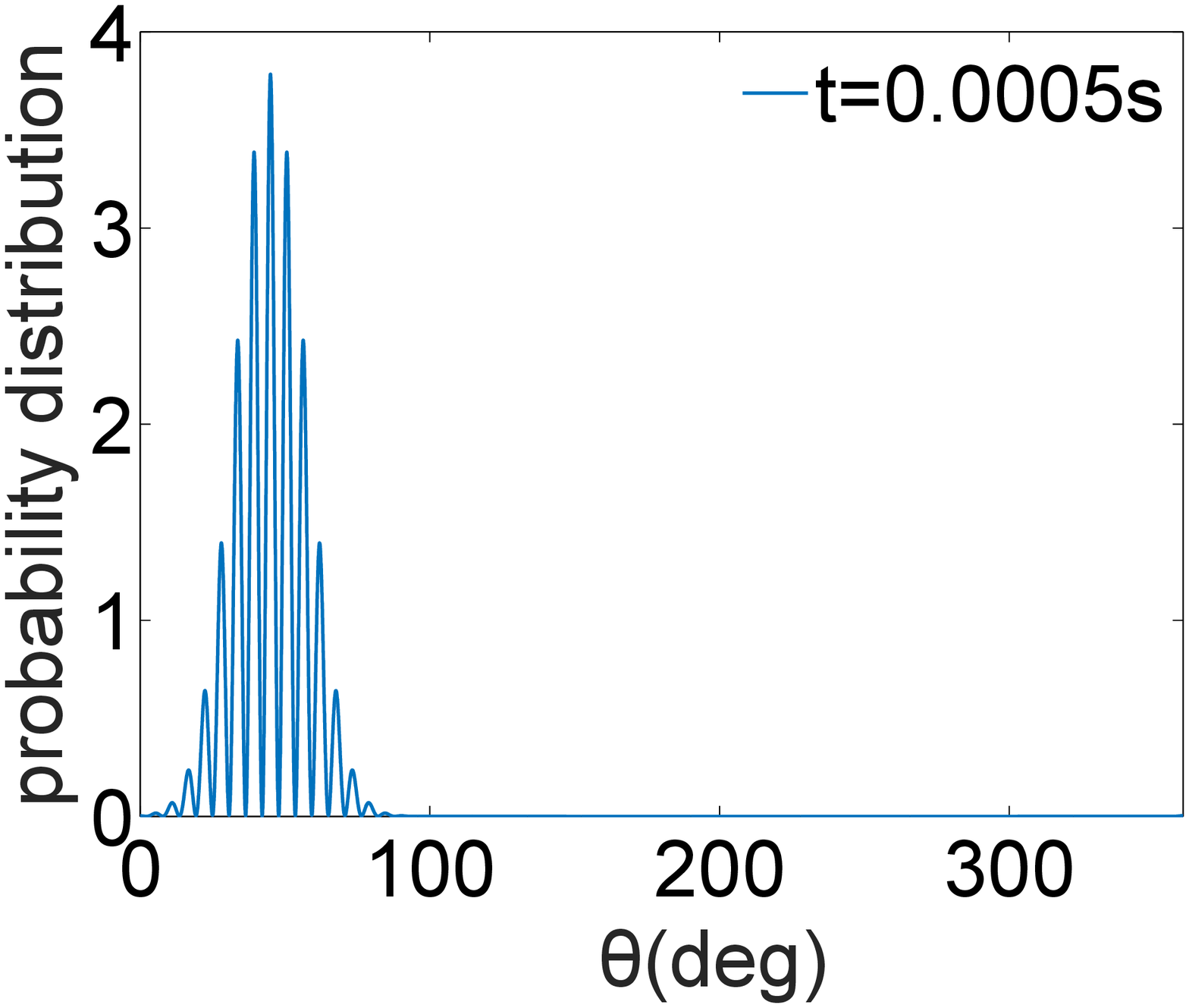}
    \includegraphics[width=.3\textwidth]{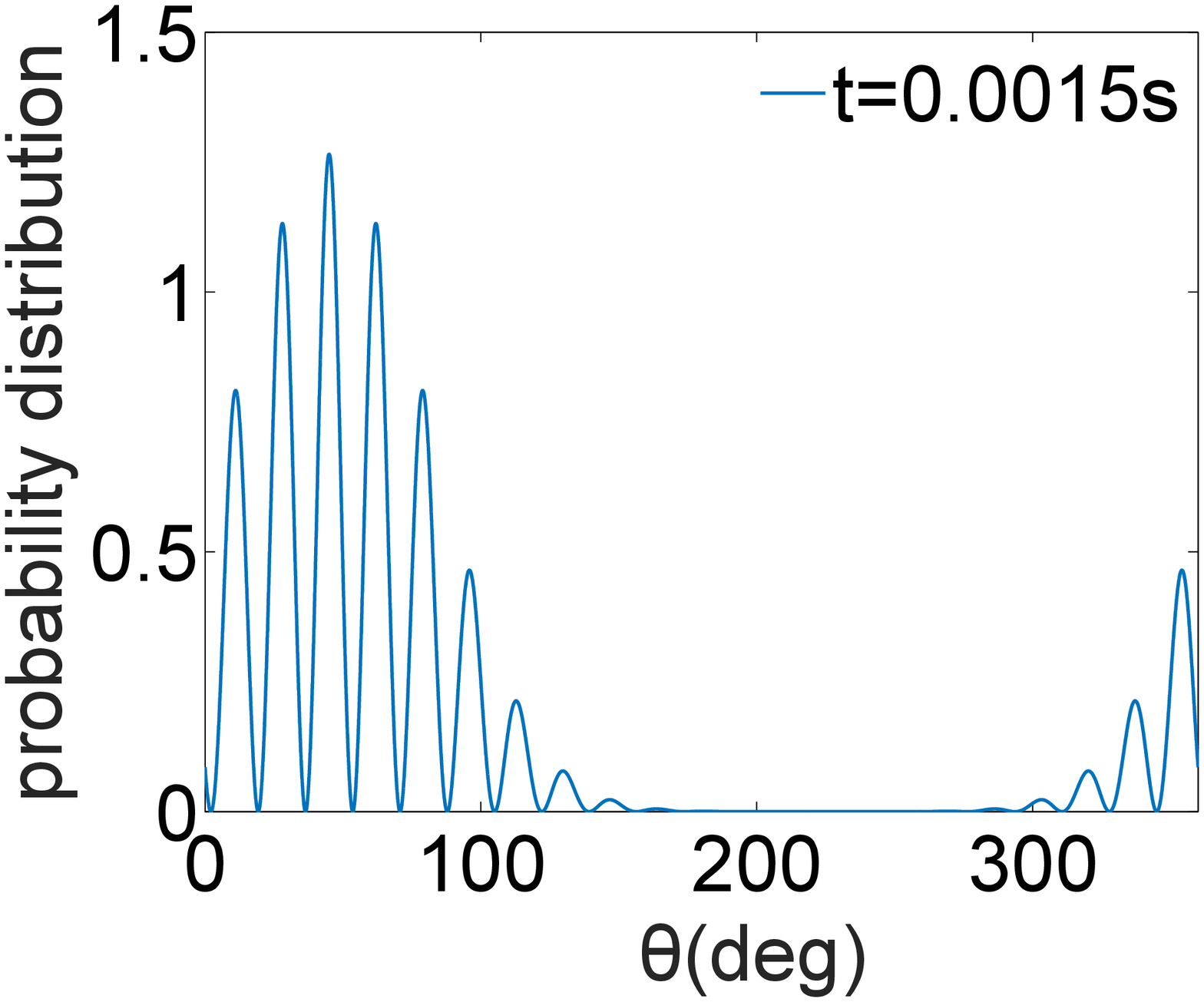}
    \includegraphics[width=.3\textwidth]{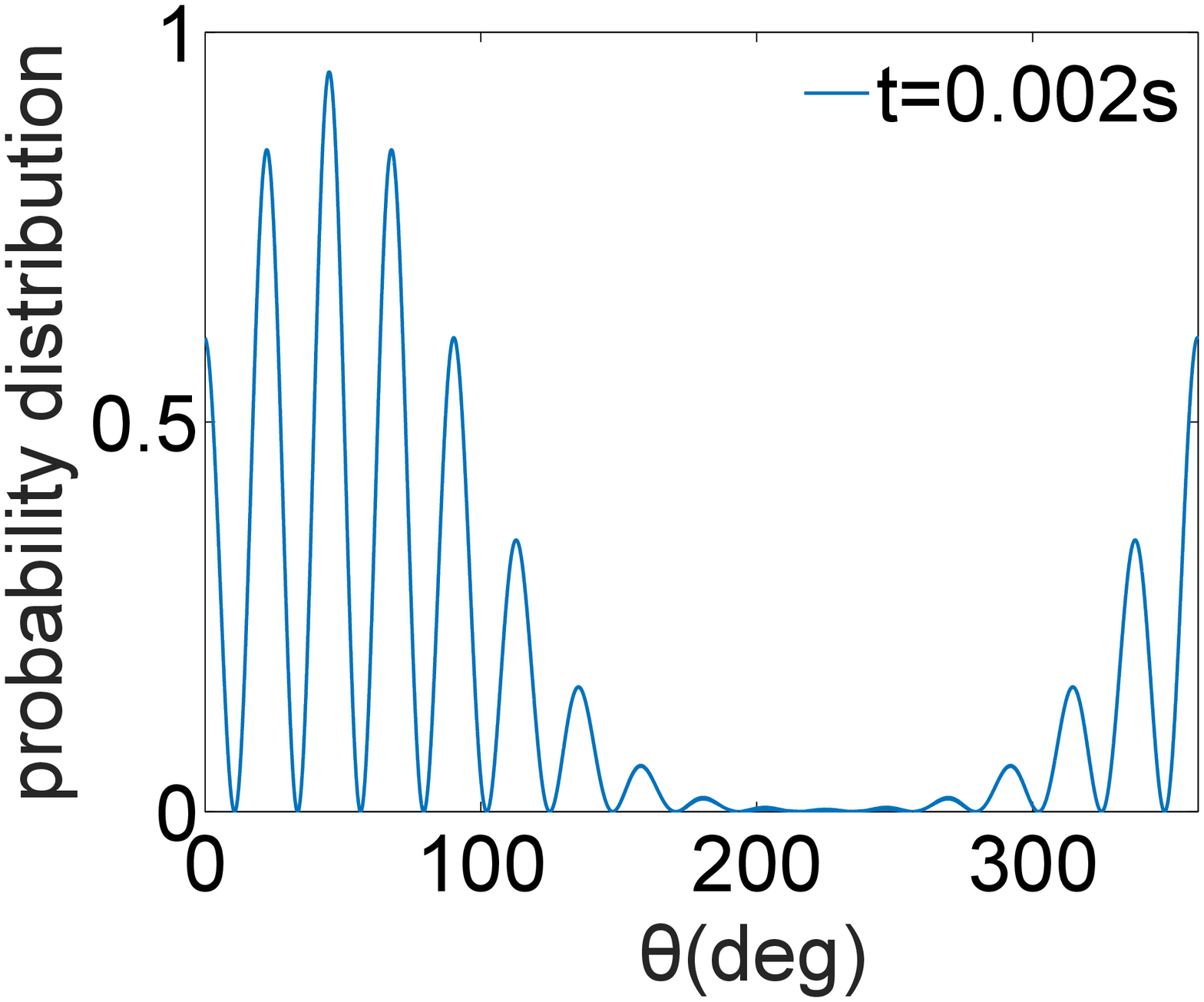}
    \includegraphics[width=.3\textwidth]{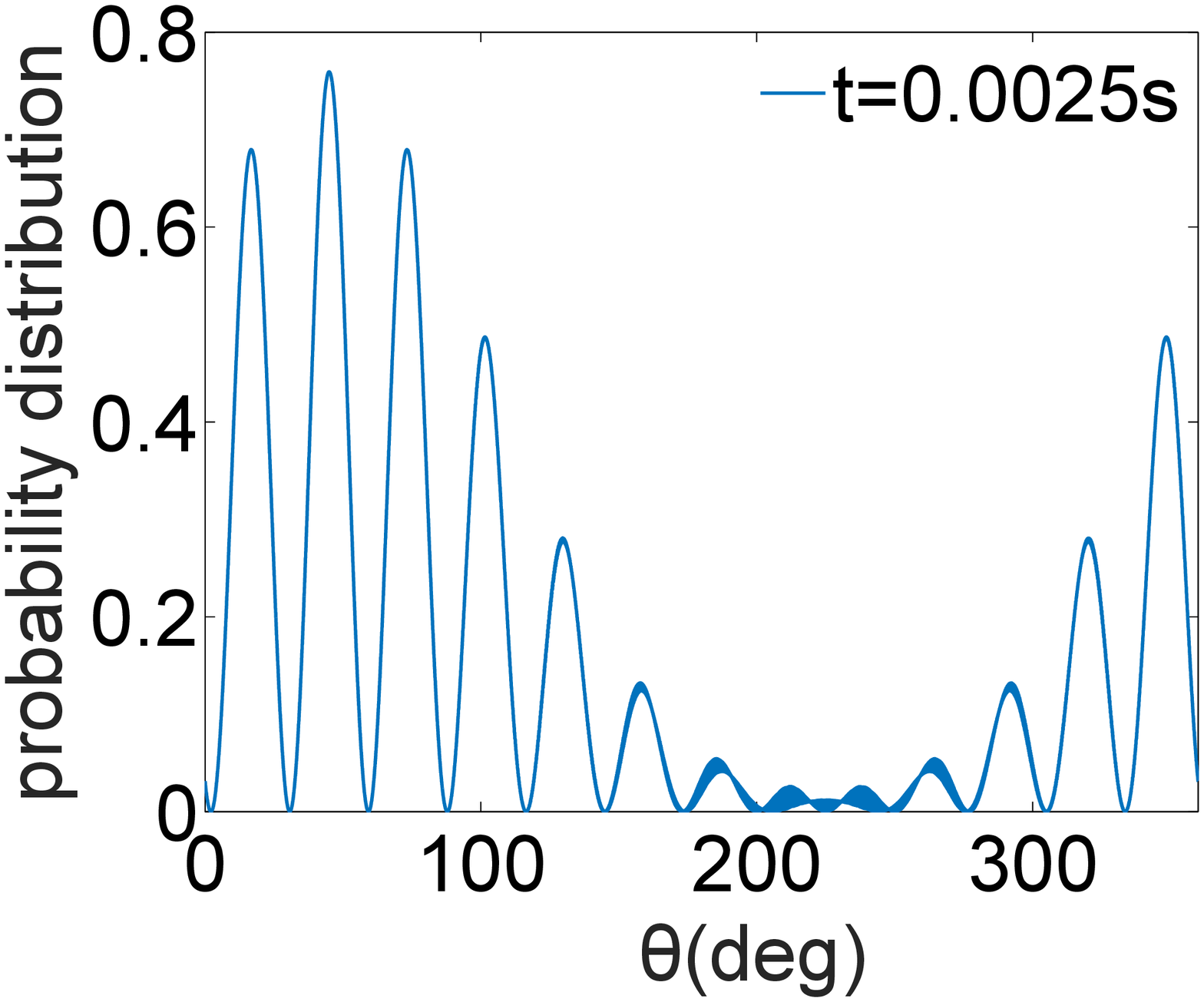}
    \includegraphics[width=.3\textwidth]{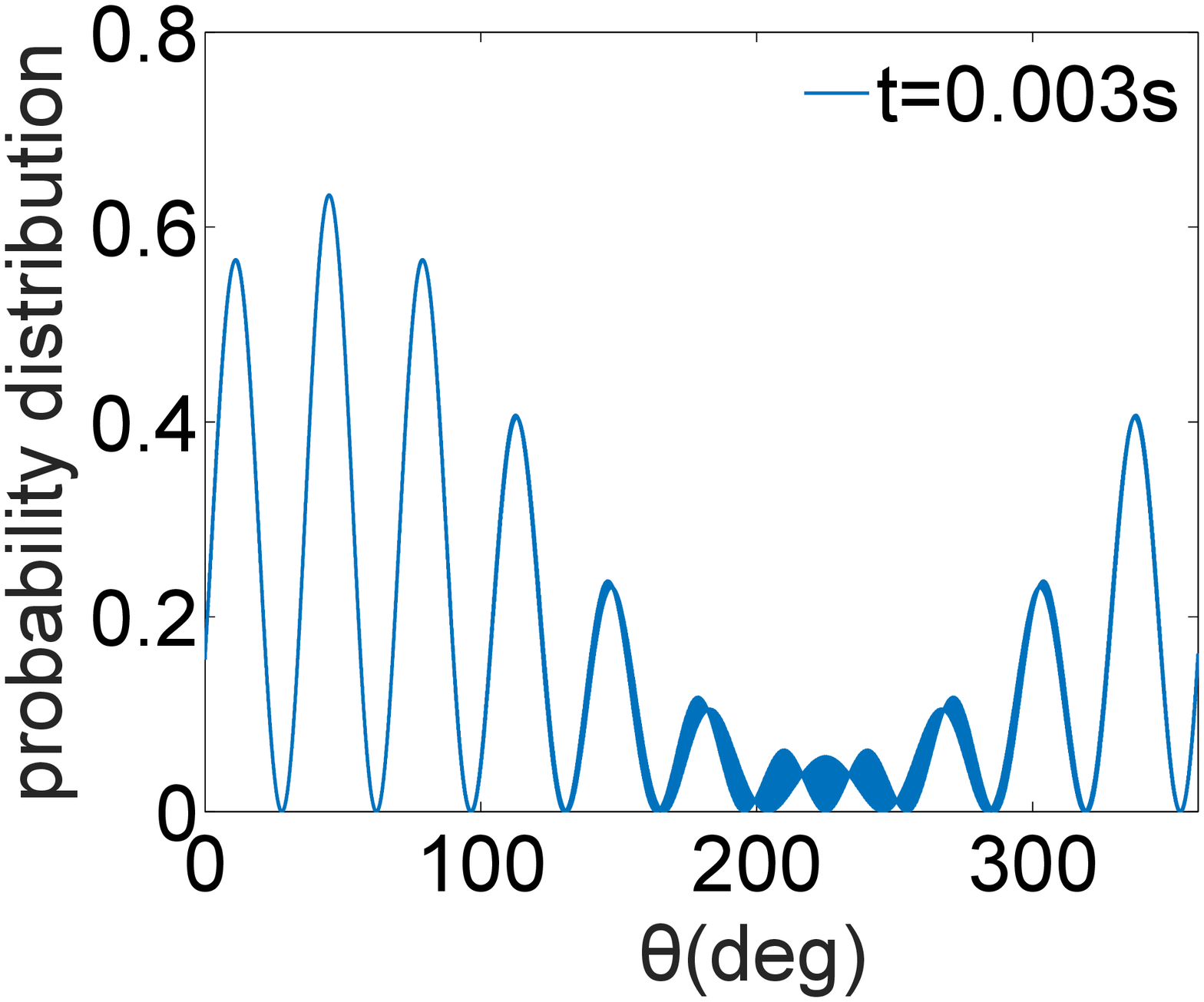}
  \caption{Time evolution of interference fringe for a nano-diamond with long semi-axis $40\ \rm{nm}$ and short semi-axis $20\ \rm{nm}$. The external magnetic field is $0.05\ \rm{T}$. The coupling strength is $g_N=2\pi\times331\ \rm{kHz}$ with the equilibrium orientation $\theta_0=\pi/4$. The trapping frequency is $\omega'_{\theta}=2\pi\times50\ \rm{kHz}$.
  When $t=0$, two original peaks are there, but no interference.  At time $t=0.0005$ s, two peak start to interference, but the distribution of fringes is still limited. At time
  $t=0.001$ s,  wavefunction continues to expand, but they still do not occupy the whole circle. At time $t=0.0025$ s, the wavefunction starts to occupy the whole circle. At around $225^{\circ}$, the ``unwanted'' meeting happens. The fringes become unclear around that region. But other regions are not affected yet. When $t=0.003$ s, wide regions are affected by the superposition effect on a circle. We can estimate that the spacing between two fringes is around $30^{\circ}$, which is large enough to be detected experimentally.
  } \label{whole}
\end{figure*}

We have used the assumption above to extend the integral interval. Then we can use two formulae of integral for Gaussian distributions: $
\int_{-\infty}^{+\infty}e^{-kx^2}dx=\sqrt{\frac{\pi}{k}}$ for $\Re(k)>0$, $\int_{-\infty-i\kappa}^{+\infty-i\kappa}e^{-\frac{z^2}{2}}dz=\int_{-\infty}^{+\infty}e^{-\frac{z^2}{2}}dz=\sqrt{2\pi}$ for arbitrary real number $\kappa$. The equation above can finally be simplified to
\begin{equation}
A_{m}^{+}=\sqrt[4]{\frac{1}{\pi I\omega'_{\theta}}}e^{-\frac{m^2}{2I\omega'_{\theta}}}e^{-im\Big(\theta_0+\sqrt{\frac{2}{I\omega'_{\theta}}}
\frac{g_N}{\omega'_{\theta}}\Big)}
\end{equation}
So we get the expansion of $\varphi_{+}(\theta)$.
\begin{equation}
\label{plus}
\varphi_{+}(\theta)=\sqrt[4]{\frac{1}{\pi I\omega'_{\theta}}}\sum_{m=-\infty}^{+\infty}e^{-\frac{m^2}{2I\omega'_{\theta}}}\frac{1}{\sqrt{2\pi}}e^{im\Big(\theta-\big(\theta_0+\sqrt{\frac{2}{I\omega'_{\theta}}}
\frac{g_N}{\omega'_{\theta}}\big)\Big)}
\end{equation}
Similarly,
\begin{equation}
\label{minus}
\varphi_{-}(\theta)=\sqrt[4]{\frac{1}{\pi I\omega'_{\theta}}}\sum_{m=-\infty}^{+\infty}e^{-\frac{m^2}{2I\omega'_{\theta}}}\frac{1}{\sqrt{2\pi}}e^{im\Big(\theta-\big(\theta_0-\sqrt{\frac{2}{I\omega'_{\theta}}}
\frac{g_N}{\omega'_{\theta}}\big)\Big)}
\end{equation}

For every fixed $m$, the time evolution factor is $e^{-iE_mt}$, $E_m=\frac{m^2}{2I}$. So the time evolution of the superposition state becomes
\begin{equation}
\begin{aligned}
\psi_0(\theta,t)=&\sqrt[4]{\frac{1}{16\pi^3I\omega'_{\theta}}}\sum_{m=-\infty}^{+\infty}\Big(e^{im\big(\theta-(\theta_0+\sqrt{\frac{2}{I\omega'_{\theta}}}
\frac{g_N}{\omega'_{\theta}})\big)}\\
&+e^{im\big(\theta-(\theta_0-\sqrt{\frac{2}{I\omega'_{\theta}}}
\frac{g_N}{\omega'_{\theta}})\big)}\Big)
e^{-\frac{m^2}{2I\omega'_{\theta}}}e^{-i\frac{m^2}{2I}t}
\end{aligned}
\end{equation}

The probability distribution $\psi_0^{*}(\theta,t)\psi_0(\theta,t)$ should form an interference pattern around the $2\pi$ circle. As the state is not the Hamiltonian's eigenstate, the probability distribution will change with time. So the interference fringe will evolve with time, too. But to deal with this expression, we need to truncate the summation of $m$ at some finite value. This can be determined from Equation~\eqref{plus} (or~\eqref{minus}). Each term in the summation has a Gaussian part, $e^{-\frac{m^2}{2I\omega'_{\theta}}}$. As a property of Gaussian distribution, if $m$ is larger than $3\times\sqrt{2I\omega'_{\theta}}$, the contribution of $e^{-\frac{m^2}{2I\omega'_{\theta}}}$ will be only around $10^{-4}$. So this can be a criterium to truncate the summation.

Now we can discuss an example and show the fringe numerically (Fig.~\ref{fringe} in the main text). Consider a nano-diamond with long semi-axis $a=40\ \rm{nm}$ and short semi-axis $b=20\ \rm{nm}$. If the external magnetic field is taken as $B=0.05\ \rm{T}$, the maximum coupling strength will be $g_B=2\pi\times3.31\times10^5\ \rm{Hz}$, and the corresponding equilibrium angle is $\theta_0=\pi/4$. We relax the angular frequency of the trap to $\omega'_{\theta}=2\pi\times50000\ \rm{Hz}$.

When the angular separation of the two Gaussian states is the largest, we remove the trap, and take this time as $t=0$. At this time, the center orientations of the two peak are $45^{\circ}-1.01^{\circ}$, $45^{\circ}+1.01^{\circ}$. The width of the distribution is very small (as can be seen from Fig.~\ref{whole}(a)). So our approximation above is valid. The maximum of $m$ should be larger than $2000$, we take the limit as $3000$ here.

The whole evolution process of the interference fringe is shown in Fig~\ref{whole}.


\begin{thebibliography}{99}

\bibitem{ASP2014}  M. Aspelmeyer, T. J. Kippenberg, and F. Marquardt, Reviews of Modern Physics \textbf{86}, 1391 (2014).

\bibitem{PZ2012}  M. Poot and  H. S. J. van der Zant, Physics Reports \textbf{511}, 273 (2012).

\bibitem{Chen2013} Y. Chen, Journal of Physics B: Atomic, Molecular and Optical Physics \textbf{46}, 104001 (2013).

\bibitem{Hammerer2009} K. Hammerer, M. Wallquist, C. Genes, M. Ludwig, F. Marquardt, P. Treutlein, P. Zoller, J. Ye, and H. J. Kimble, Phys. Rev. Lett. \textbf{103}, 063005 (2009).

\bibitem{Connell2010} A. D. O¡¯Connell, M. Hofheinz, M. Ansmann, R. C. Bialczak, M. Lenander, E. Lucero, M. Neeley, D. Sank, H. Wang, and M. Weides, Nature \textbf{464}, 697 (2010).

\bibitem{Yin2015} Z. Q. Yin, W. L. Yang, L. Sun, and L. M. Duan, Phys. Rev. A \textbf{91}, 012333 (2015).

\bibitem{Rabl2009} P. Rabl, P. Cappellaro, M. V. Gurudev Dutt, L. Jiang, J. R. Maze, and M. D. Lukin, Physical Review B \textbf{79}, 041302 (2009).

\bibitem{Arcizet2011} O. Arcizet, V. Jacques, A. Siria, P. Poncharal, P. Vincent, and S. Seidelin, Nature Physics \textbf{7}, 879 (2011).

\bibitem{Yin2015a} Z. Yin, N. Zhao, and T. Li, Science China Physics, Mechanics \& Astronomy \textbf{58}, 1 (2015).

\bibitem{NVreview2013} M. W. Doherty,  N. B. Manson, P. Delaney,  F. Jelezkod, J. Wrachtrupe, L. C. L. Hollenberg, Physics Reports, \textbf{528} 1 (2013).

\bibitem{Balasubramanian2009} G. Balasubramanian, P. Neumann, D. Twitchen, M. Markham, R. Kolesov, N. Mizuochi, J. Isoya, J. Achard, J. Beck, and J. Tissler, Nature materials \textbf{8}, 383 (2009).

\bibitem{Kucsko2013}  G. Kucsko, P.C. Maurer, N.Y. Yao, M. Kubo, H.J. Noh, P.K. Lo, H. Park, M.D. Lukin, Nature \textbf{500}, 54 (2013).

\bibitem{Shi2015} F. Shi, Q. Zhang, P. Wang, H. Sun, J. Wang, X. Rong, M. Chen, C. Ju, F. Reinhard, H. Chen, J. Wrachtrup, J. Wang, and J. Du, Science \textbf{347}, 1135 (2015).

\bibitem{Yao2012}   N. Y. Yao, L. Jiang, A. V. Gorshkov, P. C. Maurer, G. Giedke, J. I. Cirac, M. D. Lukin, Nature Communications \textbf{3}, 800 (2012).

\bibitem{Zu2014}  C. Zu, W.-B. Wang, L. He, W.-G. Zhang, C.-Y. Dai, F. Wang, L.-M. Duan. Nature \textbf{514}, 72 (2014).

\bibitem{Cai2013} J. Cai, A. Retzker, F. Jelezko, and M. B. Plenio, Nature Physics \textbf{9}, 168 (2013).

\bibitem{Yin2013}  Z. Q. Yin, T. Li, X. Zhang, and L. M. Duan, Phys. Rev. A \textbf{88}, 033614 (2013).

\bibitem{Li2011} T. Li, S. Kheifets, and M. G. Raizen. Nature Phys. \textbf{7}, 527 (2011)

\bibitem{Romero2010} O. Romero-Isart, M. L. Juan, R. Quidant, and J. I. Cirac, New Journal of Physics \textbf{12}, 033015 (2010).

\bibitem{Chang2010} D. E. Chang, C. A. Regal, S. B. Papp, D. J. Wilson, J. Ye, O. Painter, H. J. Kimble, and P. Zoller, Proc Natl Acad Sci U S A, \textbf{107}, 1005 (2010).

\bibitem{Hoang2016} T. M. Hoang, Y. Ma, J. Ahn, J. Bang, F. Robicheaux, Z. Q. Yin, and T. Li, Phys. Rev. Lett. \textbf{117}, 123604 (2016).

\bibitem{Stickler2016} B. A. Stickler, S. Nimmrichter, L. Martinetz, S. Kuhn, M. Arndt, and K. Hornberger, Phys. Rev. A \textbf{94}, 033818 (2016).

\bibitem{trappedion} T. Delord, L. Nicolas, L. Schwab,  and G. H\'{e}tet, New J. Phys. \textbf{19}, 033031 (2017).

\bibitem{Lipkin1965} H. J. Lipkin, N. Meshkov, and A. Glick, Nuclear Physics \textbf{62}, 188 (1965).

\bibitem{Meshkov1965} N. Meshkov, A. Glick, and H. Lipkin, Nuclear Physics \textbf{62}, 199 (1965).

\bibitem{Glick1965} A. Glick, H. Lipkin, and N. Meshkov, Nuclear Physics \textbf{62}, 211 (1965).

\bibitem{Milburn1997} G. J. Milburn, J. Corney, E. M. Wright, and D. F. Walls, Phys. Rev. A \textbf{55}, 4318 (1997).

\bibitem{Gang2009} G. Chen, J-Q. Liang, and S. Jia, Optics express \textbf{17}, 19682 (2009).

\bibitem{Keeling2010} J. Keeling, M. J. Bhaseen, and B. D. Simons, Phys. Rev. Lett. \textbf{105}, 043001 (2010).

\bibitem{Opatrny2015} T. Opatrny, M. Kolar, and K. K. Das, Phys. Rev. A \textbf{91}, 053612 (2015).

\bibitem{Ortiz2005} G. Ortiz, R. Somma, J. Dukelsky, S. Rombouts, Nucl. Phys. B, \textbf{707}, 421 (2005).

\bibitem{Latorre2005} J. I. Latorre, R. Orus, E. Rico, and J. Vidal, Phys. Rev. A \textbf{71}, 064101 (2005).

\bibitem{Reslen2005}  J. Reslen, L. Quiroga and N. F. Johnson, Europhysics Letters, \textbf{69}, 8 (2005).

\bibitem{Baumann2010} K. Baumann, C. Guerlin,  F. Brennecke, and T. Esslinger, Nature \textbf{464}, 1301 (2010).

\bibitem{Hamner2014} C. Hamner, C. Qu, Y. Zhang, J. Chang, M. Gong, C. Zhang, and P. Engels, Nat. Commun. \textbf{5}, 4023 (2014).

\bibitem{Morrison2008} S. Morrison and A. S. Parkins, Phys. Rev. Lett. \textbf{100}, 040403 (2008).

\bibitem{Zibold2010} T. Zibold, E. Nicklas, C. Gross, and M. K. Oberthaler, Phys. Rev. Lett. \textbf{105}, 204101 (2010).

\bibitem{Albiez05} M. Albiez, R. Gati, J. F\"{o}lling, S. Hunsmann, M. Cristiani, and M. K. Oberthaler, Phys. Rev. Lett. \textbf{95}, 010402 (2005).

\bibitem{Kuhn2016}  S. Kuhn, A. Kosloff, B. A. Stickler, F. Patolsky, K. Hornberger, M. Arndt, J. Millen, Optica \textbf{4},356 (2017).

\bibitem{Geiselmann2013} M. Geiselmann, M. L. Juan, J. Renger, J. M. Say, L. J. Brown, F. Javier G. de Abajo, F. Koppens, and R. Quidant, Nat. Nanotechnology  \textbf{8}, 175 (2013).

\bibitem{Maclaurin2012}  D. Maclaurin, M. W. Doherty, L. C. L. Hollenberg, and A. M. Martin, Phys. Rev. Lett. \textbf{108}, 240403 (2012).

\bibitem{Knowles2014} H. S. Knowles et al., Nature Materials \textbf{13}, 21 (2014).

\bibitem{Andrich2014} P. Andrich, B. J. Alemán, J. C. Lee, K. Ohno, de las Casas, Charles F, F. J. Heremans, E. L. Hu, and D. D. Awschalom, Nano letters \textbf{14}, 4959 (2014).

\bibitem{Bermudez2011} A. Bermudez, F. Jelezko, M. B. Plenio, and A. Retzker, Phys. Rev. Lett. \textbf{107}, 150503 (2011).

\bibitem{James2007} D. James and J. Jerke, Can. J. Phys. \textbf{85}, 625 (2007).

\bibitem{Goldman2015} N. Goldman and J. Dalibard, Phys. Rev. X \textbf{4}, 031027 (2014).

\bibitem{Bukov2015} M. Bukov, L. D'Alessio, and A. Polkovnikov, Advances in Physics, \textbf{64}, 139 (2015).

\bibitem{Wei2015} B.-B. Wei, C. Burk, J. Wrachtrup,  and R.-B. Liu, EPJ Quantum Technology \textbf{2}, 18 (2015).

\bibitem{Cheng2016} R. Cheng, X. Wu, and D. Xiao, arXiv:1611.00100 (2016).

\bibitem{Neukirch2015} L. P. Neukirch, E. von Haartman, J. M. Rosenholm, and A. N. Vamivakas, Nat. Phot. 9, 653 (2015).

\bibitem{Hoang2016a} T. M. Hoang, J. Ahn, J. Bang, and T. Li, Nat. Commun. 7, 12250 (2016).


\bibitem{Dusuel2005} S. Dusuel and J. Vidal, Phys. Rev. B \textbf{71}, 224420 (2005).

\bibitem{Vidal2004} J. Vidal, G. Palacios, and R. Mosseri, Phys. Rev. A \textbf{69}, 022107 (2004).


\bibitem{Botet1983} R. Botet and R. Jullien, Phys. Rev. B \textbf{28}, 3955 (1983).

\bibitem{Marquardt2007} F. Marquardt, J. P. Chen, A. A. Clerk, and S. M. Girvin, Phys. Rev. Lett. \textbf{99}, 093902 (2007).

\bibitem{Wilson2007} I. Wilson-Rae, N. Nooshi, W. Zwerger, and T. J. Kippenberg, Phys. Rev. Lett. \textbf{99}, 093901 (2007).


\bibitem{Sakurai2011} J. J. Sakurai and J. Napolitano, Modern quantum mechanics (Addison-Wesley, 2011).



\end{thebibliography}
\end{document}